\documentclass[usenatbib]{mn2e}

\bibpunct{(}{)}{;}{a}{}{,}

\newcommand{\ltsima}{$\buildrel < \over \sim$}
\newcommand{\lsim}{\lower.5ex\hbox{\ltsima}}
\newcommand{\gtsima}{$\buildrel > \over \sim$}
\newcommand{\gsim}{\lower.5ex\hbox{\gtsima}}

\usepackage{graphicx}
\usepackage{amssymb}
\usepackage{epsfig}
\usepackage{natbib}
\def\ion#1#2{#1$\;${\small\rm\@Roman{#2}}\relax}

\title[Sk160/SMC X-1]{A detailed study of the modulation of the optical light from Sk160/SMC X-1}

\author[M.J. Coe et al.]{M. J. Coe$^{1}$,  R. Angus$^{1}$, J. A. Orosz$^{2}$ \& A. Udalski$^{3}$, \\
$^{1}$ Physics and Astronomy, University of Southampton, SO17 1BJ, UK\\
$^{2}$ Department of Astronomy, San Diego State University, San Diego, CA 92182, USA \\
$^{3}$ Warsaw University Observatory, Aleje Ujazdowskie 4, 00-478 Warsaw, Poland \\
}

\begin{document}

\date{9 April 2012}

\pagerange{\pageref{firstpage}--\pageref{lastpage}} \pubyear{2002}

\maketitle

\label{firstpage}

\begin{abstract}

{Eight years of optical photometry from OGLE-III are presented of the optical counterpart to the High Mass X-ray Binary system, SMC X-1. The optical data provide the best view to date of the modulation of the light from the system at the binary period of 3.9d. In addition, it is shown for the first time that the light is also modulated at the superorbital period - a period associated with the precessing of a warped accretion disk around the neutron star partner. The implications for the sources of the various components of the optical light in this system are discussed.

}

\end{abstract}


\section{Introduction and background}

  A survey of the literature reveals $\sim$200 High Mass X-ray Binaries (HMXBs) known in our Galaxy and the Magellanic Clouds (Liu et al., 2005, 2006).  Of these, there is a substantial population of HMXBs in the Small Magellanic Cloud (SMC) comparable in number to the Galactic population (Coe et al., 2009). Of the spectrally classified HMXBs in the Galaxy there are 21 supergiant systems and 28 Be star systems in the current literature. However, unlike that Galactic population, all except one of the SMC HMXBs are Be star systems. The one exception is the supergiant HMXB system SMC X-1 which is known to be unusual in having a rather low mass neutron star in the binary partnership (Val Baker, Norton, \& Quaintrell 2005).

  SMC X-1 is a binary system consisting of a neutron star, with pulse period 0.71s (Leong et al. 1971), and the BOI super-giant companion Sk160 (Webster et al. 1972; Liller 1973). The system has an orbital period of 3.89d and a variable superorbital period, (ranging from 40 to 70 days) with an average value of 55.8d (Trowbridge et al. 2007). A warped accretion disk which partly obscures the neutron star when periodically passing in front of it is thought to be responsible for the superorbital modulations (Katz, 1973), and the variation in the superorbital period is explained by Clarkson et al. (2003) as due to multiple warp modes in the disk.

  In this paper the relationship between the X-ray modulations and the optical light is explored by making use of high-quality OGLE-III photometric monitoring over eight years. Detailed modelling of the binary modulation permits explicit confirmation of the unusually low mass of the neutron star. In addition, for the first time, evidence is presented for the presence of an optical modulation at the super-orbital period of the system.

\section{Optical data}

\begin{figure}
\includegraphics[angle=-0,width=80mm]{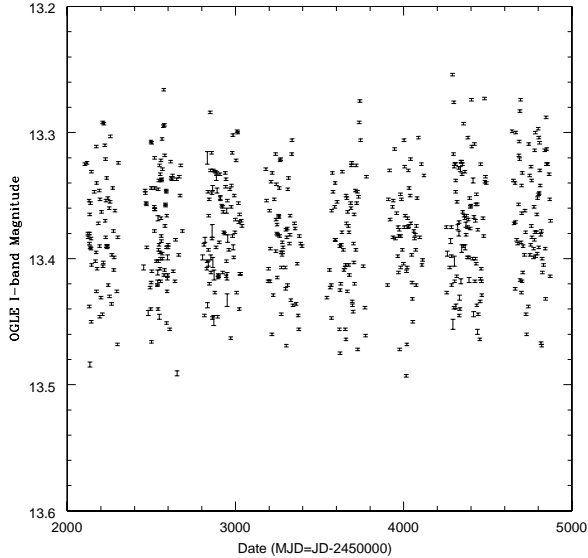}
\caption{OGLE-III I-band photometry of the optical counterpart within the SMC X-1 system.}
\label{fig:figog}
\end{figure}

Optical photometric data of the SMC X-1 system were obtained from the OGLE-III project (Udalski, 2003). The I-band data were collected approximately every night during the SMC observing season for eight years (April 2001 to June 2009). These data are illustrated in Figure~\ref{fig:figog}. It is immediately apparent from inspecting this figure that there is an underlying variability of the data with a range of $\pm0.1$ magnitudes.

\section{X-ray data}

\begin{figure}
\includegraphics[angle=-0,width=80mm]{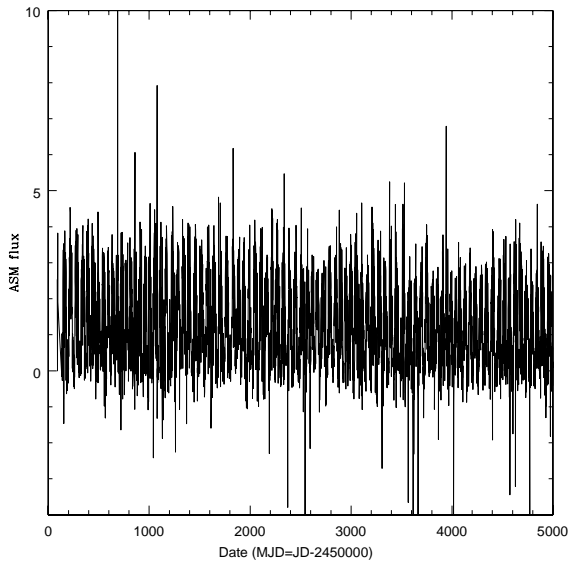}
\caption{RXTE/ASM dwell-by-dwell X-ray data for SMC X-1. The source is clearly detected for over 13 years.}
\label{fig:figasm}
\end{figure}

The ASM website presents the X-ray data  in two different formats:  daily averages - reliable data with small errors; and 'dwell by dwell' - a 90s exposure data point for every ninety minute long orbit which, individually have larger errors. After experimenting with both formats the 'dwell by dwell' data proved more appropriate for this work due to the increased sample size. These data are shown in Figure~\ref{fig:figasm}.

\section{Data analysis}

\subsection{The orbital period}

\begin{figure}
\includegraphics[angle=-0,width=80mm]{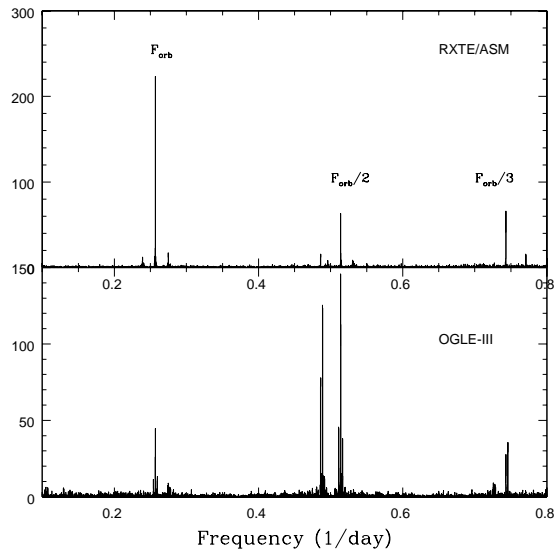}
\caption{Lomb-Scargle power spectra for SMC X-1 over the period range 1.25 -- 10d. The power spectrum from the X-ray data is shown in the upper panel and that from the optical data in lower panel. The orbital period and its harmonics are indicated in the top panel.}
\label{fig:figls}
\end{figure}

Lomb-Scargle  periodograms (Lomb 1976; Scargle 1982) were produced from the ASM and OGLE data. The orbital period is a prominent peak in both the X-ray and optical power spectra, see Figure~\ref{fig:figls}. . The second highest peak in the optical data is the product of the window function - a periodicity introduced by the daily sampling method of the OGLE data. The binary period found in the ASM data was 3.89229090$\pm$0.0000003 days - in good agreement with the value presented in Wojdowski et al. (1998) - see below.

\begin{figure}
\includegraphics[angle=-0,width=80mm]{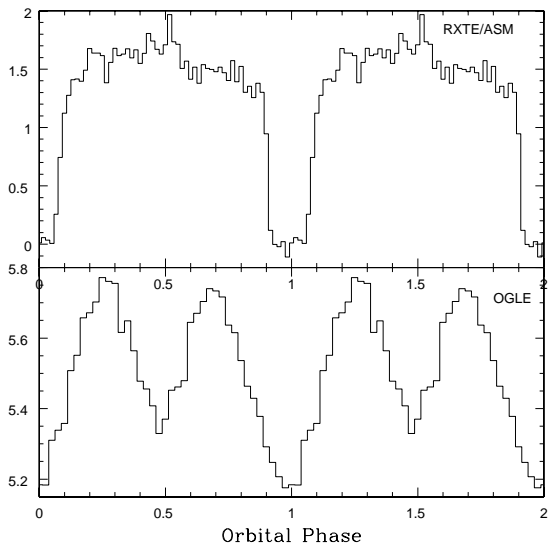}
\caption{Data folded at the binary period - see text for details of the binary period. X-ray data (upper panel) from RXTE/ASM in units of ASM counts/s. Optical data (lower panel) from OGLE-III is in flux
units of $10^{-15}erg ~cm^{-2}s^{-1}\AA^{-1}$ and folded with the same orbital parameters as the X-ray data.}
\label{fig:figfold}
\end{figure}

The next step was to fold both optical and X-ray data sets on the orbital period. However in the case of SMC X-1 the orbital period is not constant: it is decaying (Wojdowski et al. 1998) and the time of central eclipse drifts over the years. The centre time of the Nth eclipse as defined by Wodjowski et al. is given by $t_N = a_o + a_{1}N + a_{2}N^{2}$, where $a_0$ defines the initial zero-phase point: 2442836.18278(20) (days), $a_{1}$ is the orbital period: 3.89229090(43) (days), $a_{2} = -6.953(28)\times10^{-8}$ and N is the number of orbits since $a_0$. The additional 'squared' term compensates for the orbital decay. A barycenter correction was also applied to the data and the resulting folded X-ray and optical (I band) light curves are shown in
Figure~\ref{fig:figfold}.

The neutron star eclipse is the dominating feature in the upper panel of Figure~\ref{fig:figfold}, the RXTE/ASM folded light curve, manifesting between phases 0.85 and 1.15.

Some considerable effort was made to reproduce the details shown in the equivalent figure (Fig. 1) in Trowbridge et al (2007). Data from the same time period were used and data points were removed for which the RXTE/ASM solution gave $\chi^{2}\ge1.5$ (as was done by Trowbridge et al.), and then values $\ge1.3$. Various combinations of the RXTE/ASM 'A', 'B' and 'C' band data (C is the hardest band, A the softest) were tried, but without success. Unfortunately, we failed to replicate their light curve and, in particular, were unable to see the mid-eclipse X-ray 'bounce-back' feature that they report in their work.

It was noted that if we included all RXTE/ASM data in our light curve then the scatter on each point was greater than if only data from the time period reported in Trowbridge et al (2007) were used. This is probably due to the gradual deterioration of the RXTE/ASM over the years. Therefore only X-ray data from this time period were used in this work.

Since the X-ray flux of SMC X-1 varies by almost 100\% over the course of each superorbital cycle, in an attempt to improve the quality of our folded light curve only data near the times of maximum flux were used i.e. when the superorbital modulations were at the peak of their cycles. Upon producing this new folded light curve, however, the base level sections of the light curve were improved but the eclipse suffered from the reduced sample size (which was now three times smaller than it was originally).  Other combinations of superorbitally selected data were selected in an attempt to further refine the folded light curve, but ultimately decided against - not wanting to over-manipulate the data.
 	
The optical light curve shown in the lower panel of Figure~\ref{fig:figfold} exhibits the classical ellipsoidal variations of a binary system.  What is not clear, however, is the asymmetrical nature of the light curve: one would expect modulations to be equal, peaks and dips to reach the same maximum and minimum flux value. This is has been interpreted successfully by Rawls et al (2011) in terms of X-ray heating, ellipsoidal variations and a possible accretion disk. This fit is revisited and addressed further in the Discussion section of the paper.

\begin{figure}
\includegraphics[angle=-0,width=80mm]{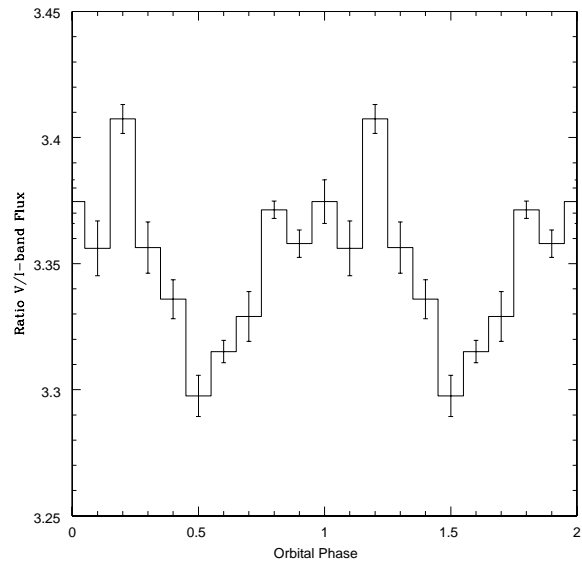}
\caption{Ratio of the flux in the OGLE-III V-band to that in the I-band as a function of binary phase.}
\label{fig:ogcol}
\end{figure}

 Following on from the shape of the folded I band light curve, it was decided to include the less-sampled V band data from OGLE-III, with the intention of creating a folded colour (V-I) phase diagram. In an attempt to maximise the signal-to-noise ratio the possibility was explored of including more V band data from Penfold and Warren (1975) and Paradijs (1977). However, these extra data were eventually discarded as they only degraded the much superior quality of the OGLE-III data. This meant sacrificing sample size, hence a lower phase resolution colour light curve of only fifteen phase bins. This was produced and this is shown in Figure~\ref{fig:ogcol}. This colour graph reveals an increase in temperature at the time of eclipse (phase 0.0) and a decrease in temperature as the X-ray source passes in front of the star.


\subsection{The superorbital period}

\begin{figure}
\includegraphics[angle=-0,width=90mm]{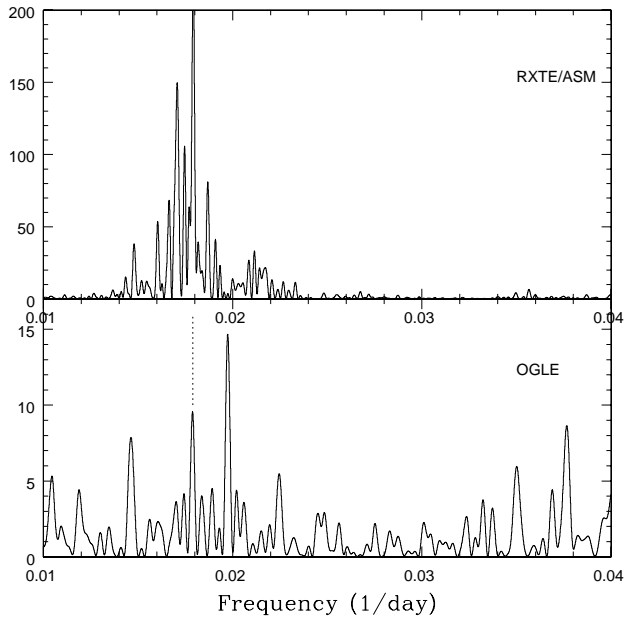}
\caption{Lomb-Scargle periodograms of the X-ray (upper panel) and optical (lower panel) data sets showing the region around the super-orbital period. The vertical dotted line in the lower panel indicates the peak position from the upper panel and represents a period of 55.8d.}
\label{fig:sopls}
\end{figure}

Once again Lomb-Scargle periodograms were produced from the RXTE/ASM and OGLE-III I-band data, this time covering the period range reported for the superorbital period - see Figure~\ref{fig:sopls}. In the X-ray data a strong but complicated peak is revealed at 55.8 days (the average superorbital  period) with the power of $\sim$200.  This same period is also present in the OGLE data, albeit with something of a lower Lomb-Scargle power - a modest $\sim$10.

A larger peak at 50.7d can be seen to the right of the optical superorbital spike in the lower panel of Figure~\ref{fig:sopls}. The origin of this feature is unclear.  A randomising test was done on the optical data which returned L-S power values of 12.4, 15.0 and 17.4 at 99\%, 99.9\% and 99.99\% levels of confidence respectively. So this peak is significant at the $\sim99.9\%$ level. One possibility is stellar non-radial pulsation with a period of 0.978 days beating with the one day sampling to produce this feature. There are several examples of this effect reported in Bird et al.(2012) in their study of the OGLE data of another similar class of HMXBs - Be/X-ray binary systems.

Nonetheless, the possibility that this period could be related to the superorbital period was explored by splitting the OGLE data into three equal length sections and producing periodograms of each. If this period wandered around following the original superorbital period, then Lomb-Scargle of each third should look different.  While the true superorbital period does indeed vary, this new peak remains at the same approximate period throughout - offering support to the non-radial pulsation model.

Despite the low power of the secondary peak at 55.8d, however, we can be confident that some of the optical light is being modulated on the same superorbital period as the X-ray radiation.

\begin{figure}
\includegraphics[angle=-0,width=80mm]{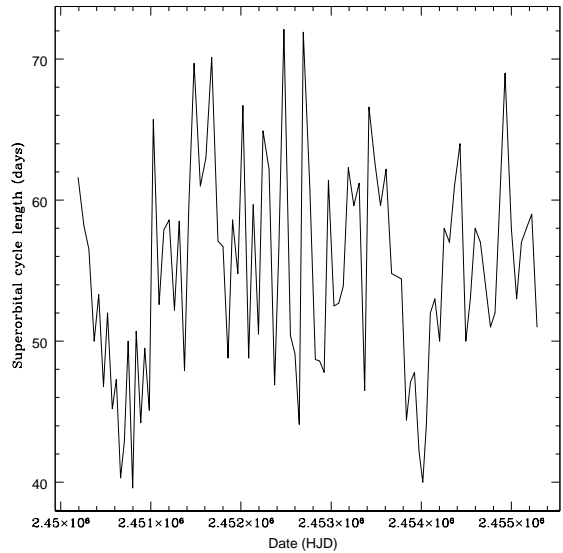}
\caption{The length of the superorbital cycle over many years.}
\label{fig:sop}
\end{figure}

Since the superorbital period shows considerable variation (see, for example, Clarkson et al., 2003), folding upon the average period was not appropriate. The RXTE/ASM data were therefore folded using Table 2 in Trowbridge et al. (2007) which logs the times of superorbital zero-phase. This table continues up to the date of publication of their work, and so had to be extended up to make it possible to include all the optical data presented here. The RXTE/ASM X-ray data were smoothed using a five point moving average and the zero-phase points were estimated by eye. An updated edition of Fig.5 from Trowbridge et al. (2007) showing the time evolution of the superorbital cycle length is presented here in Figure~\ref{fig:sop}.

\begin{figure}
\includegraphics[angle=-0,width=80mm]{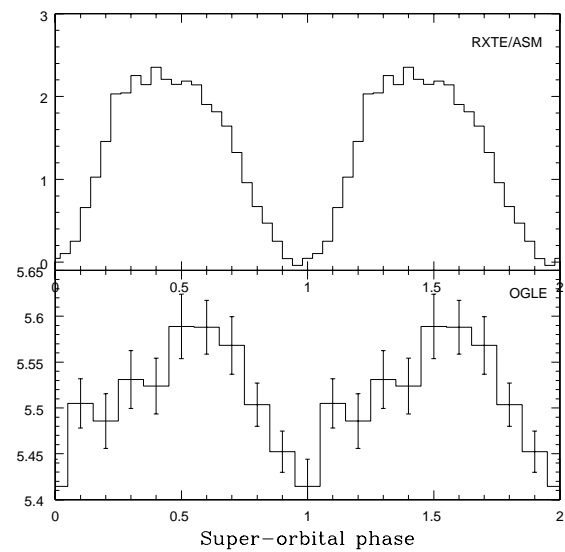}
\caption{Data folded at the super-orbital period - see text for details. X-ray data (upper panel) from RXTE/ASM in units of ASM counts/s. Optical data (lower panel) from OGLE-III I-band in flux
units of $10^{-15}erg ~cm^{-2}s^{-1}\AA^{-1}$.}
\label{fig:sopfold}
\end{figure}

The folded RXTE/ASM and OGLE light curves produced as a result of folding the data at this varying superorbital period are shown in Figure~\ref{fig:sopfold}. The ASM light curve shows a clean shape (as would be expected from its L-S periodogram) with $\sim$100\% modulation. The OGLE light curve, on the other hand, is not so strongly modulated and only varies by $\sim$3\%. $\chi^{2}$ tests on the binned OGLE data were carried out: firstly using the ASM light curve as a model; then a straight line centered on the mean value. We calculated reduced $\chi^{2}$ values of 1.3 and 111.8 respectively, providing strong support for a correlated optical and X-ray superorbital modulation. This is the first time such an effect has been noted.

\section{Discussion \& modelling}

Looking first at the binary modulation of the I-band data (Figure~\ref{fig:figfold}) we note the extra depth of the dip at phase 0.0 (the time of the eclipse of the X-ray source) compared to phase 0.5 in this folded light curve. The difference in the depths of the dips is $\sim$2--3\%. Therefore one explanation for the discrepancy between the two minimima is that there is a small component of I-band light coming from the the accretion disk itself which is, of course, eclipsed at the same phase as the X-ray source. All the rest of the I-band modulation is due to the asymmetry in the shape of the supergiant star. The fact that the visible band temperature is warmer during X-ray eclipse (see Figure~\ref{fig:ogcol}) adds further information, suggesting that the the source of the visible light in the accretion disk is somewhat cooler than the surface of the B0I star - which is typically 30,000K.

From the superorbital modulation shown in Figure~\ref{fig:sopfold} it is possible to determine that $\sim$3\% of the I-band light is modulated at the superorbital period. Since the nature of the superorbital modulation in the X-ray is attributed to the precession of a warped accretion disk (Clarkson et al., 2003), it is therefore very likely that the source of visible light is also being modulated in the same manner. If this is the case, and the optical variation mirrors the 100\% modulation seen in the X-ray profile, then one can make another estimate of the fraction of the light coming directly from the accretion disk, which is $\sim$3\%. Therefore both estimates of the fraction of I-band light arising from the accretion disk are in broad agreement at $\sim 2-3\%$. Though the exact origin of the optical superorbital modulation is still unclear, it worth noting that Rajoelimanana, Charles \& Udalski (2011) suggest, based upon other OGLE data, that such long-term modulation may be present in many more SMC sources and hence may be more widespread than previously reported. In their paper they present an apparent correlation between orbital and superorbital periods in 17 systems (their Figure 45) and suggest that the two periods could be linked by the size of the circumstellar disks around the Be stars and the light therefrom. Interestingly the two periods seen in SMC X-1 do not place it comfortably upon an extension of their plot. This may be related to the suggestion that supergiant systems do not form circumstellar disks but only exhibit radial wind outflows.

Rawls et al. (2010) recently redetermined the system parameters for six eclipsing high-mass X-ray binaries, including SMC X-1.  The  Eclipsing Light Curve (ELC)  code of Orosz \& Hauschildt (2000) was used to model all of the observational constraints in a self-consistent way. In the case of SMC X-1, the observational constraints include a $V$-band light curve from van Paradijs \& Kuiper (1984), the semi-amplitude of the O-star's radial velocity curve $K_{rm opt}$, the X-ray eclipse semiduration $\theta_e$, the orbital period $P$, and the projected semimajor axis of the pulsar's orbit $a_X\sin i$ (see van der Meer et al. 2007 and the cited references for a tabulation of these values).  Rawls et al. showed that the fit to the optical light curve was improved when an accretion disk was included in the model.  In particular, the accretion partially eclipses the O-star, increasing the depth of the minimum near phase 0 (the inferior conjunction of the pulsar).

\begin{figure}
\includegraphics[angle=-0,width=80mm]{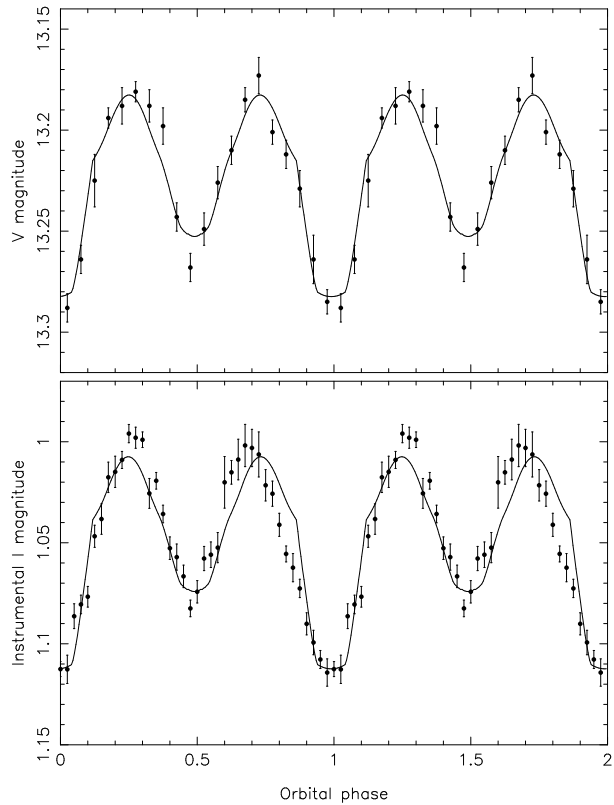}
\caption{V \& I-band data phase-folded at the orbital period and compared to synthetic lightcurves - see text for details.}
\label{fig:jo1}
\end{figure}

We extended the analysis performed by Rawls et al. (2010) by including the mean phase-averaged OGLE light curve.  When the model was extended to the $I$-band it provided a very good fit to the mean OGLE light curve.  An optimization using a Monte Carlo Markov Chain resulted in $M_{ns} = 1.145 +/- 0.074 M_{\odot}$, $M_{opt}=17.34 +/- 0.90 M_{\odot}$, and an inclination of $i=63.2 +/- 2.1 ^\circ$.  Figure~\ref{fig:jo1} shows the phased light curves and the best-fitting models.
There are some small systematic differences near the quadrature phases in the fit to the OGLE light curve, possibly caused by some asymmetry in the accretion disk.  In spite of these small deviations, the inclination of the binary is well constrained, leading to well defined masses for the components.

The neutron star mass is a little higher than that found in Rawls et al. (2010), however, it confirms that it is very much on the low side for systems that have determined neutron star masses (see, for example, the sample in Rawls et al, 2011 and references therein).

\section{Acknowledgements}

 The OGLE project has received funding from the European Research Council under the European Community's Seventh Framework Programme (FP7/2007-2013)/ERC grant agreement no. 246678 to AU.

\bsp

\label{lastpage}


\begin{thebibliography}{99}

\bibitem[]{} A. J. Bird, M. J. Coe, V. A. McBride \& A. Udalski (2012) MNRAS 423, 3663.

\bibitem[]{} Clarkson, W. I., Charles, P. A., Coe, M. J., Laycock, S., Tout, M. D., \& Wilson, C. A. 2003, MNRAS, 339, 447.



\bibitem[]{} Coe M.J. et al. 2009 The Magellanic System: Stars, Gas, and Galaxies, Proceedings of the International Astronomical Union, IAU Symposium, Volume 256, p. 367-372.



\bibitem[]{} Katz, J. I. 1973, Nature, 246, 87.

\bibitem[]{} Leong, C., Kellogg, E., Gursky, H., Tananbaum, H., \& Giacconi, R. 1971, ApJ, 170, L67.

\bibitem[]{} Liller, W. 1973, ApJ, 184, L37.

\bibitem[]{} Liu, Q. Z.; van Paradijs, J.; van den Heuvel, E. P. J. 2005, A\&A 442, 1135.

\bibitem[]{} Liu, Q. Z.; van Paradijs, J.; van den Heuvel, E. P. J. 2006, A\&A 455, 1165.





\bibitem[]{} Lomb, N. R. 1976, ApSS, 39, 447

\bibitem[]{} Orosz, J. A. \& Hauschildt, P. H. 2000, A\&A, 364, 265



\bibitem[]{} Prsa, A. \& Zwitter, T. 2005, ApJ , 628, 426.

\bibitem[]{} Rajoelimanana A.~F., Charles P.~A., Udalski A., 2011, MNRAS, 413, 1600.

\bibitem[]{} Rawls M.~L., Orosz J.~A., McClintock J.~E., Torres M.~A.~P., Bailyn C.~D.,
Buxton M.~M., 2011, ApJ, 730, 25


\bibitem[]{} Scargle, J. D. 1982, ApJ, 263, 835

\bibitem[]{} Trowbridge, S., Nowak, M. A., \& Wilms, J. 2007, ApJ, 670, 634

\bibitem[]{} van der Meer A., Kaper L., van Kerkwijk M.~H., Heemskerk M.~H.~M., van den Heuvel E.~P.~J., 2007, A\&A, 473, 523.

\bibitem[]{} van Paradijs, J., \& Kuiper, L. 1984, A\&A, 138, 71.


\bibitem[]{} Udalski, A. 2003, Acta Astron., 53, 291.

\bibitem[]{} Val Baker, A.K.F., Norton, A.J., \& Quaintrell, H., 2005, A\&A 441, 685.

\bibitem[]{} Webster, B. L., Martin, W. L., Feast, M. W. \& Andrews, P. J. 1972, Nature, 240, 183

\bibitem[]{} Wilson, R. E. \& Devinney, E. J. 1971, ApJ , 166, 605.


\bibitem[]{} Wojdowski, P., Clark, G. W., Levine, A. M., Woo, J. W. \& Zhang, S. N. 1998, ApJ, 502, 253










\end{thebibliography}
\end{document}